\documentclass[aps,prapplied,reprint,groupedaddress]{revtex4-2}
\usepackage{lineno}
\usepackage{graphicx}% Include figure files
\usepackage{subfigure}
\usepackage{float}
\usepackage{dcolumn}% Align table columns on decimal point
\usepackage{makecell}
\usepackage{bm}% bold math
\usepackage{upgreek}
\usepackage[colorlinks=true,linkcolor=blue,citecolor=blue,urlcolor=blue]{hyperref}% add hypertext capabilities
\makeatletter

\newcommand{\Rmnum}[1]{\expandafter\@slowromancap\romannumeral #1@}
\makeatother
\begin{document}
%\preprint{}

\title{An extremely bad-cavity laser}


\author{Jia Zhang$^{1}$}
\author{Tiantian Shi$^{1,2,3}$}
\email{tts@pku.edu.cn}
\author{Jianxiang Miao$^{1}$}
\author{Deshui Yu$^{4}$}
\author{Jingbiao Chen$^{1,5}$}
\affiliation{$^{1}$ State Key Laboratory of Advanced Optical Communication Systems and Networks, Institute of Quantum Electronics, School of Electronics, Peking University, Beijing 100871, China}
\affiliation{$^{2}$ School of Integrated Circuits, Peking University, Beijing 100871, China}
\affiliation{$^{3}$ National Key Laboratory of Advanced Micro and Nano Manufacture Technology, Beijing 100871, China}
\affiliation{$^{4}$ National Time Service Center, Chinese Academy of Sciences, Xi’an 710600, China}
\affiliation{$^{5}$ Hefei National Laboratory, Hefei 230088, China}

\begin{abstract}
Lasing in the bad-cavity regime has promising applications in precision measurement and frequency metrology due to the reduced sensitivity of the laser frequency to cavity length fluctuations. Thus far, relevant studies have been mainly focused on conventional cavities whose finesse  is high enough that the resonance linewidth is sufficiently narrow compared to the cavity’s free spectral range, though still in the bad-cavity regime. However, lasing output from the cavity whose finesse is close to the limit of 2 has never been experimentally accessed. Here, we demonstrate an extremely bad-cavity laser, analyze the physical mechanisms limiting cavity finesse, and report on the worst ever laser cavity with finesse reaching 2.01. The optical cavity has a reflectance close to zero and only provides a weak optical feedback. The laser power can be as high as tens of $\mu \mathrm{W}$ and the spectral linewidth reaches a few kHz, over one thousand times narrower than the gain bandwidth. In addition, the measurement of cavity pulling reveals a pulling coefficient of 0.0148, the lowest value ever achieved for a continuous wave laser. %Benefiting from the simple structure, high portability, and robustness against cavity pulling, the extremely bad-cavity laser potentially serves as a compact active optical clock that has various uses in out-of-the-laboratory environment and chip-scale metrology.
Our findings open up an unprecedentedly innovative perspective for future new ultra-stable lasers, which could possibly trigger the future discoveries in optical clocks, cavity QED, continuous wave superradiant laser, and explorations of quantum manybody physics.

\end{abstract}

\maketitle


\section{Introduction}
Lasers are one of the greatest inventions of the twentieth century and have been the most versatile tool available for scientific, industrial, and medical applications, owing to their directionality, high brightness, monochromaticity, and high degree of coherence \cite{sixtylaser}. Ultrastable lasers with ultranarrow linewidths are highly desired for precision spectroscopy and fundamental physics measurements. Recently, the ground-based observatory consisting of twin laser interferometers has enabled the detection of ripples in space-time caused by binary black hole mergers \cite{abbott2016observation}; optical clocks based on ultrastable lasers have resolved the gravitational redshift at millimeter scale \cite{bothwell2022resolving,zheng2022differential}; laser-based gyroscopes allow measurements of Earth’s rotation-induced optical path change of the order of $10^{-16}$~cm, which is one thousandth of the classical electron radius \cite{liang2017resonant,lai2020earth}; and optical whispering-gallery sensors are capable of detecting single molecules and even ions \cite{yu2021whispering}. In all these applications, it is necessary for lasers to be (pre-)stabilized to high-finesse optical reference cavities using Pound-Drever-Hall (PDH) technique \cite{drever1983laser}. Stabilizing the laser frequency to a passive reference cavity with a higher stability not only substantially suppresses the laser spectrum broadening but also improves the (short-term) laser frequency stability.

\begin{figure*}
	\includegraphics[width=\linewidth]{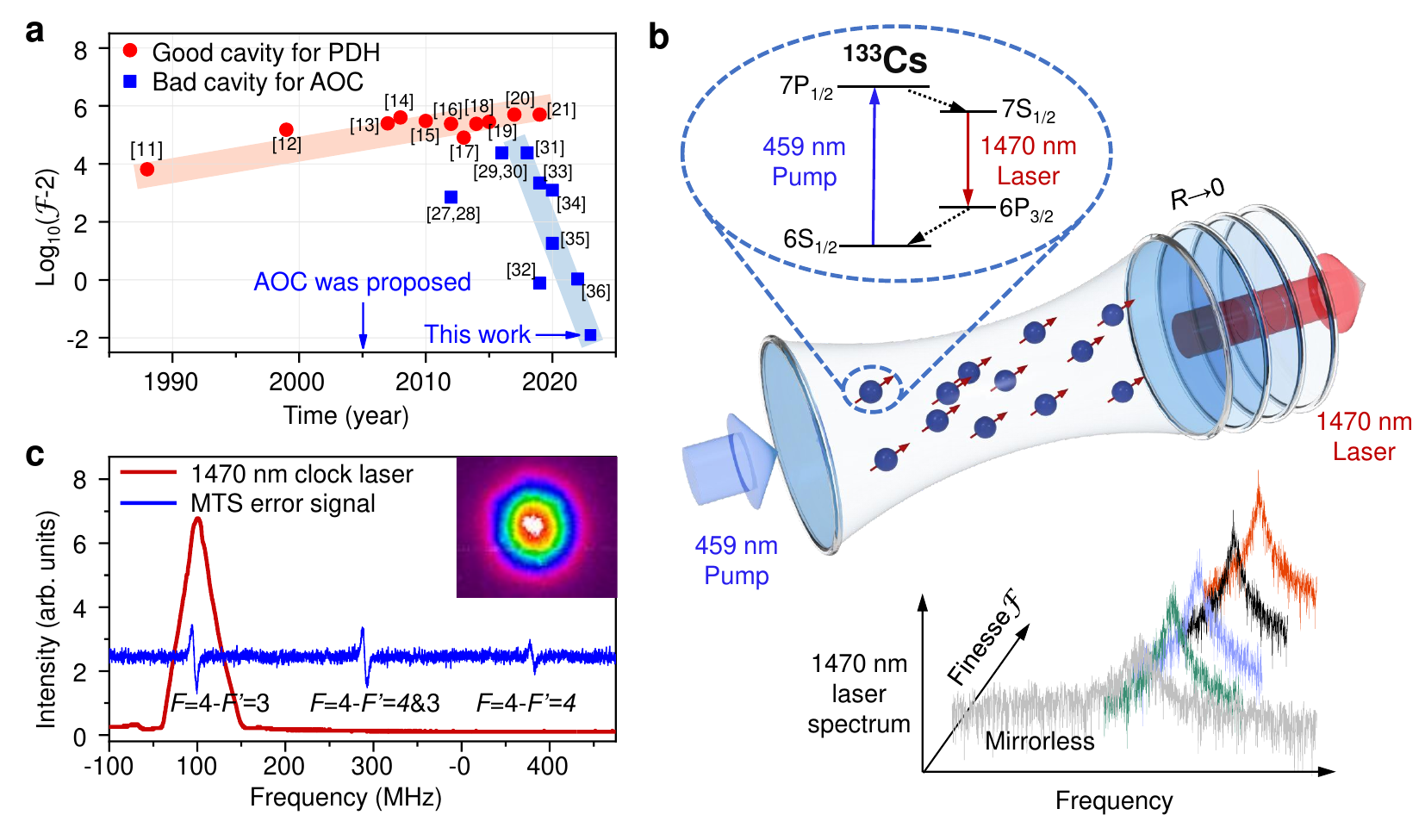}
	\caption{\label{fig1}
		\textbf{Experimental scheme.} \textbf{a.} Trend of logarithmic finesse $\mathrm{Log}_{10}(\mathcal{F}-2)$ \cite{shi2023antiresonant} over time. Red circle symbols correspond to the good-cavity finesse in Pound-Drever-Hall frequency stabilization method \cite{salomon1988laser,young1999visible,ludlow2007compact,alnis2008subhertz,jiang2010nd,kessler2012sub,chen2013frequency,bloom2014optical,hafner20158,matei20171,robinson2019crystalline} and blue square symbols denote the bad-cavity finesse in active optical clocks (AOCs) \cite{bohnet2012steady,bohnet2012relaxation,norcia2016cold,norcia2016superradiance,norcia2018frequency,Shi2019good,laske2019pulse,schaffer2020lasing,liu2020rugged,shi2022inhibited}. \textbf{b.} Schematic diagram of extremely bad-cavity laser. A 459-nm laser drives thermal Cs atoms from the ground $6\mathrm{S}_{1/2}$ to the excited $7\mathrm{P}_{1/2}$ level. The population inversion is achieved between upper $7\mathrm{S}_{1/2}$ and lower $6\mathrm{P}_{3/2}$ levels through the spontaneous emission from $7\mathrm{P}_{1/2}$ to $7\mathrm{S}_{1/2}$. An extremely bad cavity with the cavity reflectance $R = r_{1}r_{2}$ and amplitude reflection coefficients $r_{1,2}$ of cavity mirrors is used to introduce the weak optical feedback. The lasing action at the wavelength of 1470 nm occurs when the pump exceeds the optical loss. The laser spectrum is narrowed as $R$ grows, as shown in the bottom image. \textbf{c.} Lasing at 1470 nm. The blue curve denotes the dispersion-shaped modulation transfer spectroscopy (MTS) signal that is used to stabilize the 459-nm pump laser to the $6\mathrm{S}_{1/2}(F=4)$ - $7\mathrm{P}_{1/2}(F^{\prime}=3)$ hyperfine transition in Cs. The red curve shows the corresponding 1470-nm laser power. Inset: Transverse mode of the 1470-nm laser imaged on a high-sensitivity fast photodetector.
	}
\end{figure*}

Although conventional super-stabilized lasers have high application value, they operate in the good-cavity regime, where the linewidth of the laser cavity $\Gamma_{\mathrm{cavity}}$ is much narrower than the optical gain bandwidth $\Gamma_{\mathrm{gain}}$, and laser cavity-length fluctuations strongly influence the laser frequency $\nu_{0}$ through the so-called cavity pulling effect \cite{siegman1986lasers}. Therefore, it is essential for the optical reference cavity to be carefully engineered so as to ensure its excellent stability. An ultrahigh finesse $\mathcal{F}$ is of great importance to the reference cavity for sensing laser frequency fluctuations since $\mathcal{F}$ determines the resonance linewidth of the reference cavity. Although the finesse of a microwave superconducting cavity can be as high as $4.6\times10^{9}$ \cite{kuhr2007ultrahigh}, the typical $\mathcal{F}$ of an optical cavity is of the order of $10^{5}$ due to the limited reflectance of (distributed Bragg) mirrors \cite{salomon1988laser,young1999visible,ludlow2007compact,alnis2008subhertz,jiang2010nd,kessler2012sub,chen2013frequency,bloom2014optical,hafner20158,matei20171,robinson2019crystalline} (see Fig.~\ref{fig1}a), which leads to a resonance linewidth at, for example, the 10~kHz level for a cavity length of 10~cm. In addition, great efforts such as choosing crystalline silicon as cavity material, placing the cavity in a vacuum chamber, and cryogenic cooling of the cavity have to be made to maximally isolate the reference cavity from environmental perturbations (e.g., vibrations and Brownian-motion thermal noise) \cite{kessler2012sub,robinson2019crystalline}. The linewidth of a laser stabilized to such a cryogenic reference cavity can be narrowed down to 5~mHz, corresponding to a coherence time of tens of seconds \cite{matei20171}. Nevertheless, %the high complexity and huge volume of the entire laser system significantly restricts its portability and use outside the laboratory.
even for the state-of-the-art stable laser using ultrahigh finesse cavity, the inevitable cavity length thermal noise introduces a time-integrated phase drift that makes it difficult to break through the Schawlow-Townes limit \cite{schawlow1958infrared,kuppens1994quantum,yu2008laser,chen2009active}.

In contrast to the pursuit of the high finesse, a bad-cavity laser requires a cavity relaxation rate that exceeds the atomic relaxation rates by several orders of magnitude, which requires a low-finesse cavity. Indeed, it has been recognized that lasing in the bad-cavity regime, where $\Gamma_{\mathrm{cavity}}\ll\Gamma_{\mathrm{gain}}$, strongly reduces the sensitivity of the laser frequency $\nu_{0}$ to cavity length fluctuations so that $\nu_{0}$ is primarily determined by the atomic transition frequency $\nu_{a}$ \cite{yu2008laser,chen2009active}. Such bad-cavity lasers may directly serve as active optical clocks without the need of extra laser frequency stabilization to atomic transitions \cite{yu2008laser,chen2009active,bohnet2012steady,bohnet2012relaxation,norcia2016cold,norcia2016superradiance,norcia2018frequency,Shi2019good,laske2019pulse,schaffer2020lasing,liu2020rugged,shi2022inhibited,yu2023proposal,yu2023active}. In addition, the relatively long coherence time of the macroscopic polarization of active atoms substantially suppresses the laser phase noise \cite{kolobov1993role}, resulting in a laser linewidth that overcomes the usual Schawlow–Townes limit \cite{schawlow1958infrared,kuppens1994quantum,yu2008laser}. Reducing the cavity reflectance $R$ leads to the drop of the laser cavity finesse $\mathcal{F}$ and broadens the cavity linewidth $\Gamma_{\mathrm{cavity}}$, which further suppresses cavity pulling. Contrary to the pursuit of high-finesse cavity in PDH frequency stabilization, bad-cavity lasers expect as low finesse as possible, which results in two diametrically opposed trends of the cavity finesse, as shown in Fig.~\ref{fig1}a. Thus far, most laser cavities used in experiments have a reflectance higher than 0.3 and a logarithmic finesse value $\mathrm{Log}_{10}(\mathcal{F}-2)$ exceeding 0. %The resultant cavity linewidth $\Gamma_{\mathrm{cavity}}$ is sufficiently narrower than the free spectral range FSR of the cavity. 
Here, $\mathcal{F}$ approaching 2 denotes the mirrorless radiation (i.e., in the absence of the cavity) \cite{casperson1977threshold}. Nonetheless, lasing in the extremely bad-cavity regime, where $R$ is close to zero (i.e., $\mathcal{F}\sim2$) and the laser cavity only provides extremely weak optical feedback, has never been accessed. In this work, we demonstrate for the first time the extremely bad-cavity laser whose mirror reflectance is close to zero, which may lead to new research in the laser physics, detection of gravitational wave, cavity QED fields and so on. In addition to the interesting connections with many areas of fundamental science, we also anticipate that new applications and technologies will continue to emerge from the study of the physics of extremely bad-cavity laser.

\begin{figure*}
	\includegraphics[width=\linewidth]{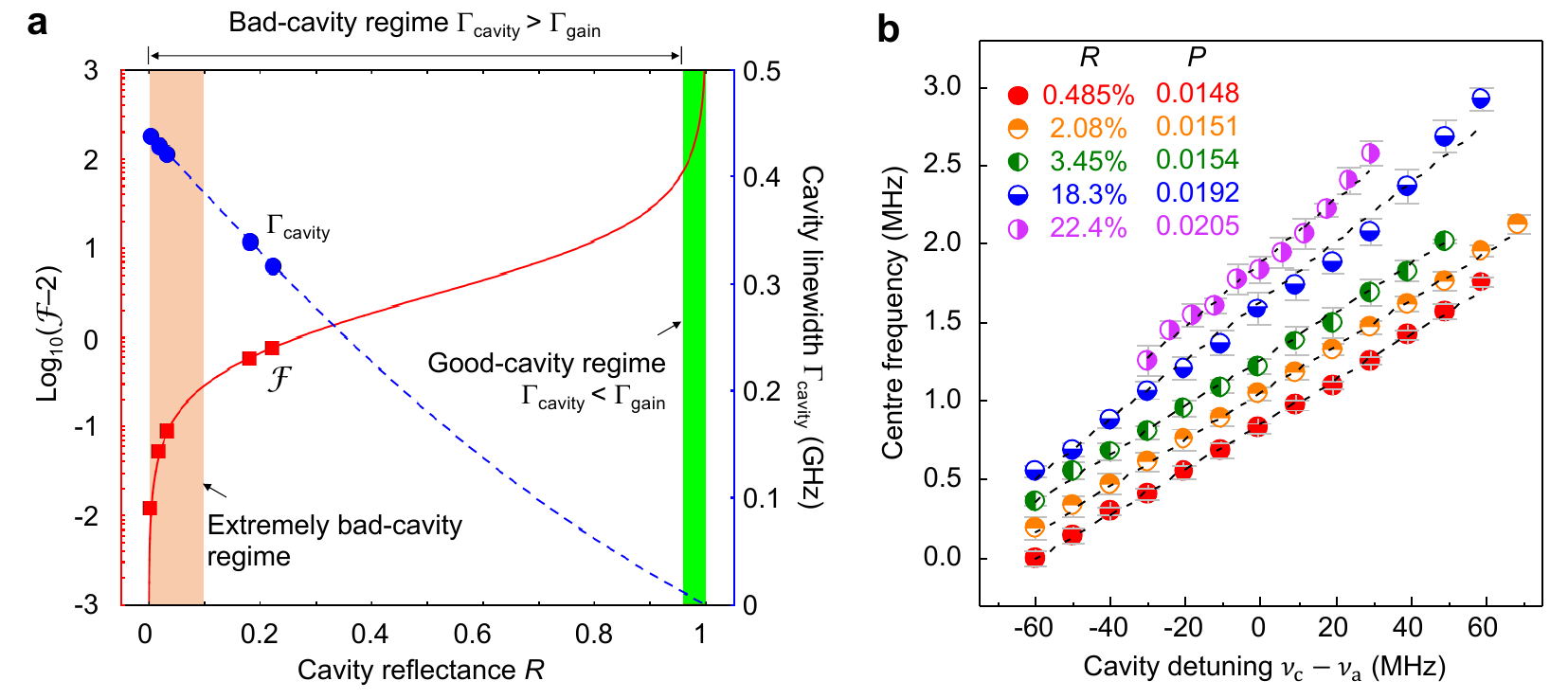}
	\caption{\label{fig2}
		\textbf{Cavity-pulling effect.} \textbf{a.} Dependence of finesse on cavity reflectance. In the extremely bad-cavity regime, the cavity reflectance $R$ is close to zero. Symbols: logarithmic cavity finesse $\mathrm{Log_{10}(\mathcal{F}-2)}$ and cavity linewidth $\Gamma_{\mathrm{cavity}}$ used in experiment. Solid line: Eq.~(\ref{eq.1}). Dashed line: $\Gamma_{\mathrm{cavity}}$ derived from Eq.~(\ref{eq.1}) with the free spectral range FSR~=~882~MHz. \textbf{b.} Sensitivity of the 1470-nm laser frequency to the detuning $\delta$ of the optical cavity from the atomic transition frequency for different cavity reflectances $R$. Symbols: central frequency of beat note between two 1470-nm lasers in the same environment. The detuning of one laser is set at zero and the detuning of the other varies. Dash-dotted lines: linear curve fitting. The slope of a curve denotes the corresponding cavity-pulling coefficient at $\delta$. For all measurements, the pump power is set at 5.2~mW and the temperature is $116.2^{\circ}\mathrm{C}$.
	}
\end{figure*}

\section{RESULTS}
\textbf{Experimental scheme.} 
Fig.~\ref{fig1}b illustrates the general setup of the experiment, where a cylindrical glass cell (diameter of 2~cm and length $l = 5~\mathrm{cm}$) of Cs atoms is placed inside an invar optical cavity (length $L = 17~\mathrm{cm}$ and free spectral range FSR = 882~MHz) whose finesse is adjustable. A home-made 459-nm external cavity diode laser (spot area $S = 1.69~\mathrm{mm}^{2}$) with an interference filter configuration is used to drive atoms in the ground $6\mathrm{S}_{1/2}$ state to the excited $7\mathrm{P}_{1/2}$ state (see the inset in Fig.~\ref{fig1}b). To suppress the frequency drift (so as to provide a continuous pump), the 459-nm pump laser is stabilized to the $6\mathrm{S}_{1/2}(F=4)$ - $7\mathrm{P}_{1/2}(F^{\prime}=3)$ hyperfine transition in Cs through the modulation transfer spectroscopy as shown in Fig.~\ref{fig1}c (for more details refer to Ref. \cite{miao2022compact}).

Atoms are accumulated in $7\mathrm{S}_{1/2}$ via the spontaneous emission decay from $7\mathrm{P}_{1/2}$ to $7\mathrm{S}_{1/2}$, leading to the population inversion on the laser $7\mathrm{S}_{1/2}$ - $6\mathrm{P}_{3/2}$ transition (wavelength of 1470~nm, frequency $\nu_{a} = 204~\mathrm{THz}$, and natural linewidth of 1.81~MHz). The lasing action occurs once the pump rate exceeds the optical loss rate of the system. Taking into account the Doppler broadening, the optical gain bandwidth is estimated to be, for example, $\Gamma_{\mathrm{gain}} = 6.39~\mathrm{MHz}$ at the pump power of 5.2~mW (see Methods). The vapor cell temperature is kept at $T = 116.2^{\circ}\mathrm{C}$ to maximize the laser output power (see Methods).

We are interested in the lasing dynamics in the extremely bad-cavity regime. The cavity finesse \cite{shi2023antiresonant}
\begin{equation}\label{eq.1}
	\mathcal{F}=\frac{\mathrm{FSR}}{\Gamma_{\mathrm{cavity}}}=\frac{\pi}{\arccos(\frac{2R}{1+R^{2}})},
\end{equation}
is normally employed to measure the full width at half maximum $\Gamma_{\mathrm{cavity}}$ of the reflection Airy distribution of the optical cavity relative to the FSR. Here, $R = r_{1}r_{2}$ denotes the cavity reflectance with the amplitude reflection coefficients $r_{1,2}$ of two cavity mirrors. In the good-cavity limit, where $R$ is close to unity, $\Gamma_{\mathrm{cavity}}$  is reduced to the traditional cavity linewidth and $\mathcal{F}\approx \frac{\pi\sqrt{R}}{1-R}$ can be as high as, for example, $5\times10^{5}$ \cite{matei20171}, which yields $\Gamma_{\mathrm{cavity}}\approx1.7~\mathrm{kHz}$, much smaller than the optical gain bandwidth $\Gamma_{\mathrm{gain}}$. Reducing $R$ decreases $\mathcal{F}$, thereby broadening $\Gamma_{\mathrm{cavity}}$. The laser system enters the bad-cavity regime when $\Gamma_{\mathrm{cavity}}$  surpasses $\Gamma_{\mathrm{gain}}$. Active optical clocks are typically operated in this regime because of the strong suppression of the cavity pulling \cite{chen2009active}. Nevertheless, lasing in the extremely bad-cavity regime, where the cavity reflectance $R$ is close to zero and the cavity only provides a weak optical feedback, has never been accessed. In this extreme situation, Eq.~(\ref{eq.1}) is simplified as $\mathcal{F}\approx2+\frac{8R}{\pi(1-4R)}$  and $\Gamma_{\mathrm{cavity}}=\frac{\mathrm{FSR}}{\mathcal{F}}\approx\frac{\mathrm{FSR}}{2}$ (see Fig.~\ref{fig2}a). It is seen that as $R$ declines to zero, $\mathcal{F}$ approaches 2, at which the optical feedback completely vanishes, i.e., mirrorless, rather than zero. In what follows, we investigate the 1470-nm lasing dynamics in five extremely bad optical cavities with $R$ ranging from 0.485\% ($\mathcal{F}=2.01$) to 22.4\% ($\mathcal{F}=2.78$) (see Table~\ref{table1} in Methods) and $\mathcal{F}\approx2$ (mirrorless) as well.

\begin{figure*}
	\includegraphics[width=\linewidth]{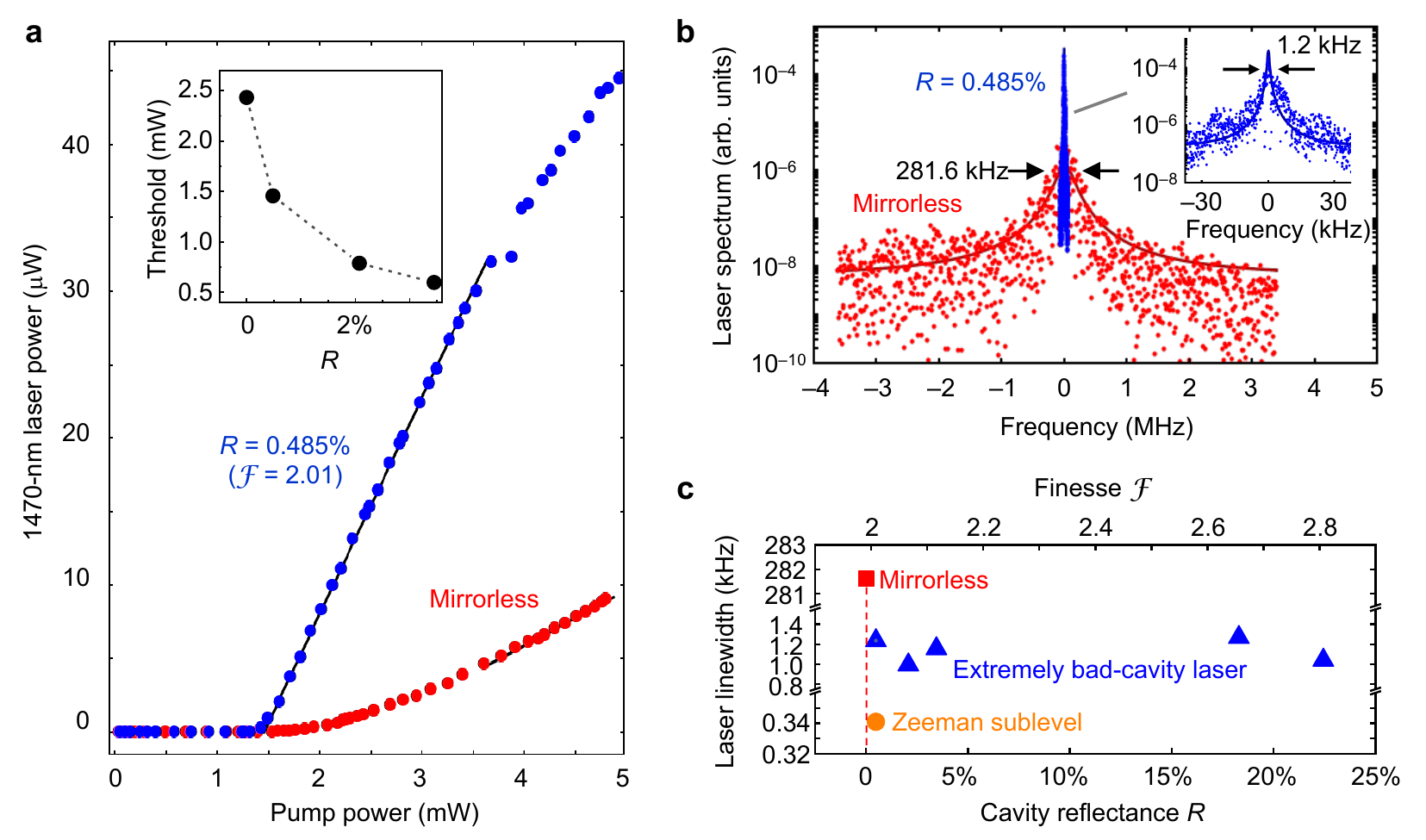}
	\caption{\label{fig3}
		\textbf{Mirrorless and extremely bad-cavity lasers.} \textbf{a.} Dependence of the 1470-nm laser power on the 459-nm pump power. Symbols: experimental results. Solid lines: linear curve fitting above the threshold. Inset: laser threshold as a function of the cavity reflectance $R$. \textbf{b.} Laser spectrum. Symbols: experimental results. The spectral linewidth is derived through the Lorentz fit (solid curves). Inset: detailed laser spectrum with $R = 0.485\%$. \textbf{c.} Laser linewidth as a function of $R$.
	}
\end{figure*}

\textbf{Cavity-pulling effect.}
According to \cite{siegman1986lasers}, the central frequency $\nu_{0}$ of a conventional laser is pulled away from the atomic transition $\nu_{\mathrm{a}}$  by an amount ($\nu_{0}-\nu_{\mathrm{a}})=P(\nu_{\mathrm{c}} - \nu_{\mathrm{a}})$ with the pulling coefficient $P$ when the cavity frequency $\nu_{\mathrm{c}}$  is detuned from $\nu_{\mathrm{a}}$. The cavity pulling effect directly maps cavity (thermal and mechanical) fluctuations onto the laser frequency, deteriorating the stability of $\nu_{0}$. As has been pointed out in \cite{kolobov1993role}, the laser phase information in the bad-cavity regime is primarily stored in the active medium’s polarization because of the atomic memory effect, and $P$ can be much smaller than unity, making the laser frequency robust against cavity fluctuations. Using cavity-mode linewidth $\Gamma_{\mathrm{cavity}}$ and atomic gain linewidth $\Gamma_{\mathrm{gain}}$, $P$ can also be expressed as $P=\frac{\Gamma_{\mathrm{cavity}}}{\Gamma_{\mathrm{cavity}}+\Gamma_{\mathrm{gain}}}$. For lasers work in bad-cavity region, where $\Gamma_{\mathrm{gain}} \ll \Gamma_{\mathrm{cavity}}$, $P \ll 1$, the effect of cavity-mode frequency variations on output laser frequency is greatly suppressed. Nevertheless, the cavity pulling in the extremely bad-cavity limit has never been explored.

We exam the cavity pulling based on our laser system. To determine the laser frequency shift $\nu_{0}-\nu_{\mathrm{a}}$, the heterodyne detection is performed on two identical 1470-nm lasers that are in the same environment (see Methods). The cavity detuning for one laser is set to be zero while the detuning $\nu_{\mathrm{c}} - \nu_{\mathrm{a}}$ for the other laser is precisely tuned within the range $-60<\nu_{\mathrm{c}} - \nu_{\mathrm{a}}<60~\mathrm{MHz}$ through adjusting the piezoelectric ceramic. The central frequency of the beat note, i.e., $\nu_{0}-\nu_{\mathrm{a}}$, can be extracted using the frequency analyzer (FA, Keysight N9000B). Measurement results for different cavity reflectivities $R$ (finesses $\mathcal{F}$) has been summarized Fig.~\ref{fig2}b, which manifests the linear relationship between $\nu_{0}-\nu_{\mathrm{a}}$ and $\nu_{\mathrm{c}} - \nu_{\mathrm{a}}$. The pulling coefficient $P$ is then derived through the linear curve fitting. 

Experimentally, $P$ can be as low as 0.0148 (as shown in Fig.~\ref{fig2}b and  Table~\ref{table1}), i.e., the cavity pulling is suppressed by almost seventy times, the strongest suppression ever achieved for a continuous-wave laser, over a wide range of about 120~MHz adjusting cavity frequency. 

\textbf{Laser power and linewidth characteristics.} 
We begin with the mirrorless radiation in the absence of the optical cavity. To avoid any reflection, vapor cell windows are oriented at Brewster's angle (see Methods). Emitting the 459-nm pump laser into thermal atoms, we observe the continuous-wave 1470-nm radiation with the fundamental mode (TEM00)  imaged on a high-sensitivity fast PIN photodetector (see Fig.~\ref{fig1}c). Analogous to conventional lasers, the 1470-nm radiation displays a threshold behavior characterized by a rapid rising power as the pump power is increased (see Fig.~\ref{fig3}a). %Spontaneously emitted photons at one end of the gain medium can be amplified along the sample and eventually leads to a large number of coherent photons emitted from the other end. 
At a high pump power, more atoms can be populated in $7\mathrm{S}_{1/2}$, not only enhancing the spontaneous emission input signal but also providing a sufficient optical gain that is essential to trigger the so-called continuous-wave amplified spontaneous emission (mirrorless lasing \cite{casperson1977threshold}).

Above the threshold, the resultant laser spectrum is also strongly narrowed down to a width $\Delta\nu = 281.6~\mathrm{kHz}$ (see Fig.~\ref{fig3}b), less than one fortieth of the optical gain bandwidth $\Gamma_{\mathrm{gain}}$. To the best of our knowledge, this is the largest spectral narrowing of the amplified spontaneous emission ever observed \cite{casperson1972spectral}. In addition, since the cavity effect vanishes in the mirrorless lasing, the laser frequency stability is completely determined by the atoms, and the corresponding Allan deviation potentially takes the form $\sigma_{y}(\tau) = \sqrt{\Delta\nu/2\pi\nu_{0}^{2}\tau} = 1.0\times10^{-12}/\sqrt{\tau}$ at the averaging time $\tau$ \cite{riehle2006frequency}. Here, $\nu_{0}$ denotes the central frequency of the 1470-nm laser.

In the presence of extremely bad cavity, the weak optical feedback is introduced and the 1470-nm laser power grows strongly as shown in Fig.~\ref{fig3}a. This is understandable because of the stimulated emission by the gain medium, the feedback is enhanced, thereby raising the output power and suppressing the lasing threshold (see the inset in Fig.~\ref{fig3}a). Remarkably, the width of the laser spectrum is dramatically suppressed down to, for example, $\Delta\nu = 1.2~\mathrm{kHz}$ with $R = 0.485\%$ ($\mathcal{F} = 2.01$), leading to a spectral narrowing factor of about 250 compared to the mirrorless lasing (see Fig.~\ref{fig3}b). It may be ascribed to the fact that the coherence of the lasing signal is enhanced by coherent photons (i.e., reflected laser portion). The resultant laser frequency stability \cite{riehle2006frequency} potentially reaches $\sigma_{y}(\tau) = 6.8\times10^{-14}/\sqrt{\tau}$. Actually, after eliminating the common-mode noise such as cavity vibrations, pump power fluctuations, and vapor cell temperature variation, the laser linewidth may be further suppressed down to $\Delta\nu = 341~\mathrm{Hz}$ (see Fig.~\ref{fig3}c and Methods). That is, the laser linewidth is extremely sensitive to the cavity reflectance within the range of $0<R<0.485\%$. However, due to the limitations in experiment, it is challenging to determine the specific sensitivity relation. As illustrated in Fig.~\ref{fig3}c, the laser linewidth $\Delta\nu$ can still be maintained around 1.2~kHz, even though $R$ approaches to 0. This is a new way of obtaining narrow linewidths in addition to good-cavity lasers using ultrahigh finesse cavities.

%One may expect that the extremely bad-cavity laser transforms to the conventional bad-cavity laser for a high enough $R$.

\begin{figure*}
	\includegraphics[width=\linewidth]{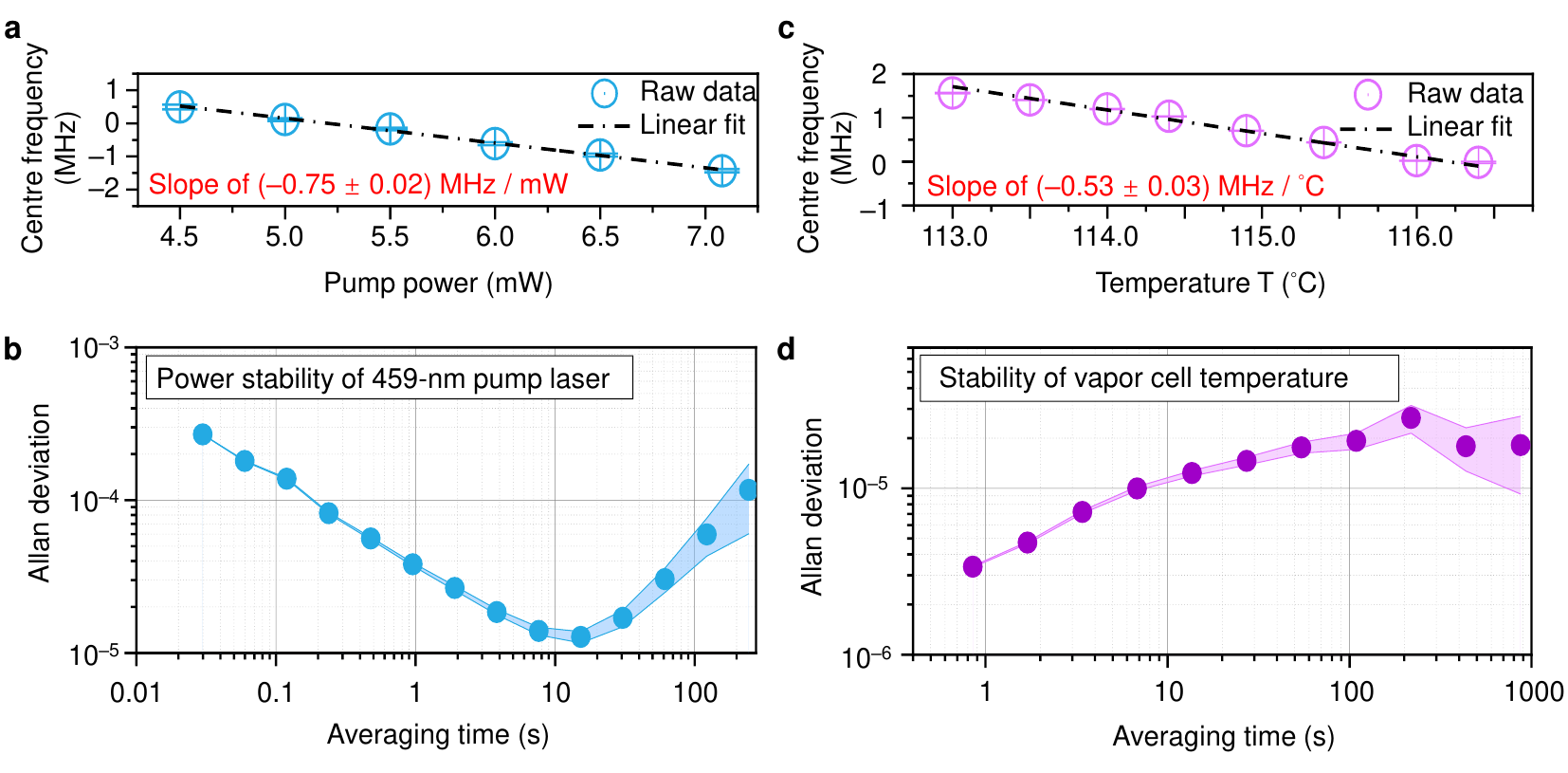}
	\caption{\label{fig4}
		\textbf{Laser frequency shift induced by technical and environmental noises.} The laser frequency shift is measured through the beat note between two independent 1470-nm lasers. \textbf{a.} Dependence of laser frequency on pump power. The pump power of one laser is set at 5.2~mW and the pump power of the other varies. The temperature of both lasers is set at $116.2^{\circ}\mathrm{C}$. Symbols: experimental results. Dash-dotted line: linear curve fitting with the resultant slope of $-750\pm20~\mathrm{kHz/mW}$. \textbf{b.} Power stability of the 459-nm pump laser. The shade area denotes the error band. \textbf{c.} Dependence of laser frequency shift on vapor cell temperature $T$. Symbols: experimental results. Dash-dotted line: linear curve fitting with the resultant slope of $-530\pm30~\mathrm{kHz}/^{\circ}\mathrm{C}$. The temperature of one laser is set at $116.2^{\circ}\mathrm{C}$ and the temperature of the other varies. \textbf{d.} Temperature stability at $116.2^{\circ}\mathrm{C}$. The shade area denotes the error band.
	}
\end{figure*}

\textbf{Technical and environmental noises.} Various noises cause the frequency shift and extra broadening of the 1470-nm laser. We focus on two main sources, pump power and vapor cell temperature fluctuations.

The frequency fluctuations of the free-running 459-nm pump laser result in the large frequency shift and spectral broadening of the 1470-nm laser. To address this issue, the pump laser is stabilized to the $6\mathrm{S}_{1/2}(F=4)$ - $7\mathrm{P}_{1/2}(F^{\prime}=3)$ hyperfine transition in Cs through the modulation transfer spectroscopy and the resultant frequency stability reaches $2.8\times10^{-13}⁄\sqrt{\tau}$ \cite{miao2022compact}. However, power fluctuations of the pump laser still strongly influence the 1470-nm laser. As shown in Fig.~\ref{fig4}a, the central frequency of the 1470-nm laser depends linearly on the pump power with a slope of $-750\pm20~\mathrm{kHz/mW}$. The power stability of the pump laser is measured to be $3.9\times10^{-5}$ at 1~s (see Fig.~\ref{fig4}b). Thus, at the typical pump power of 5.2~mW used in this work, the variation of 1470-nm laser frequency fluctuations induced by pump power fluctuations is estimated to be 152~Hz, corresponding to a fractional frequency stability of $7.5\times10^{-13}$ at 1~s of averaging. We also measure the relative intensity noise of the 459~nm laser, the results show that at 10~kHz, the RIN after frequency stabilization is increased by about 13~dB compared to unstabilized frequency. Thus, the pump power stability can be further enhanced using an acousto-optic modulator for power stabilization in subsequent experiments.

The temperature $T$ of Cs vapor cell is stabilized at $116.2^{\circ}\mathrm{C}$. Although temperature fluctuations are controlled within $0.1^{\circ}\mathrm{C}$, the vapor cell still experiences a long-term temperature drift. To evaluate the influence of temperature fluctuations on the 1470-nm laser, we measure the dependence of the 1470-nm laser frequency on $T$ over a range of $3.5^{\circ}\mathrm{C}$. It is found that the laser central frequency shifts approximately linearly with $T$ and the curve fitting gives a slope of $-530\pm30~\mathrm{kHz}/^{\circ}\mathrm{C}$ (see Fig.~\ref{fig4}c). In addition, the temperature monitoring of the vapor cell yields a temperature stability of $3.7\times10^{-6}$ at 1~s (see Fig.~\ref{fig4}d) of averaging. The variation of the 1470-nm laser frequency caused by temperature fluctuations is estimated to be 228~Hz, corresponding to a fractional frequency stability of $1.1\times10^{-12}$ the average time of 1~s. A double-layered atomic cell with vacuum heat insulation is expected to be an efficient way to suppress temperature fluctuations.

To reveal the intrinsic linewidth of the 1470-nm laser based upon thermal atoms, a weak magnetic field is used to induce energy-level splitting of the laser transition, and the beat note between 1470-nm laser beams based on different Zeeman sublevel transitions is measured. Since these Zeeman sublevel transitions share the same optical cavity, technical and environmental noises can be substantially suppressed in the beat spectrum, leading to a spectral linewidth as narrow as 482~Hz (see Methods). Assuming statistical independence and equal contribution of different 1470-nm laser beams, we infer the linewidth of 341~Hz (see Fig.~\ref{fig3}c) for each laser, implying a short-time fractional frequency stability of $3.6\times10^{-14}$.

\section{Discussion}\label{Discussion}
In this work, an extremely bad-cavity laser with cavity finesse close to the limit of 2 is demonstrated. The obtained laser power is as high as tens of $\mu\mathrm{W}$ and the laser linewidth reaches 1.2~kHz, limited by pump power and vapor cell temperature fluctuations. We also explore the suppression of cavity pulling and find a pulling coefficient of 0.0148, the lowest value ever achieved for a continuous-wave laser. Additionally, the mirrorless lasing is studied for comparison. Although the cavity pulling vanishes completely in the absence of the cavity, the mirrorless superradiance linewidth is strongly broadened by two orders of magnitude, compared to that of the extremely bad-cavity laser. That is, the laser spectrum is extremely sensitive to the weak optical feedback from the mirrorless limit to the extremely bad-cavity regime. 

In the future, to reduce the influence on the spectral linewidth and frequency stability of the 1470-nm extremely bad-cavity laser caused by the Doppler broadening and temperature fluctuations of the thermal atoms active medium, we will employ cold atoms. Recently, a one-dimensional 1-m-long sample of cold Cs atoms has been demonstrated in experiment \cite{wan2022quasi}. The resultant temperature of atoms reaches as low as 25~$\mu\mathrm{K}$ and the corresponding Doppler broadening is well below the natural linewidth of the laser (clock) transition, promising a cold-atom-based active optical clock that is operated in continuous manner and in the extremely bad-cavity regime. 

Extremely bad-cavity lasers represent a new exciting field of research that synthesizes laser physics and science of quantum pricision measurement. The physical mechanisms underlying the extremely bad-cavity lasers have never been studied before. Thus, we believe that these findings might open a completely new research area of physics such as active optical clock, cavity QED. As an important  application, the concept of an extremely bad-cavity laser offers a promising opportunity for the implementation of novelly super-stable lasers, which might overcome the practical challenges faced by ultrahigh-finesse optical cavitiy.

\section{METHODS}
\begin{figure}[H]
	\includegraphics[width=\linewidth]{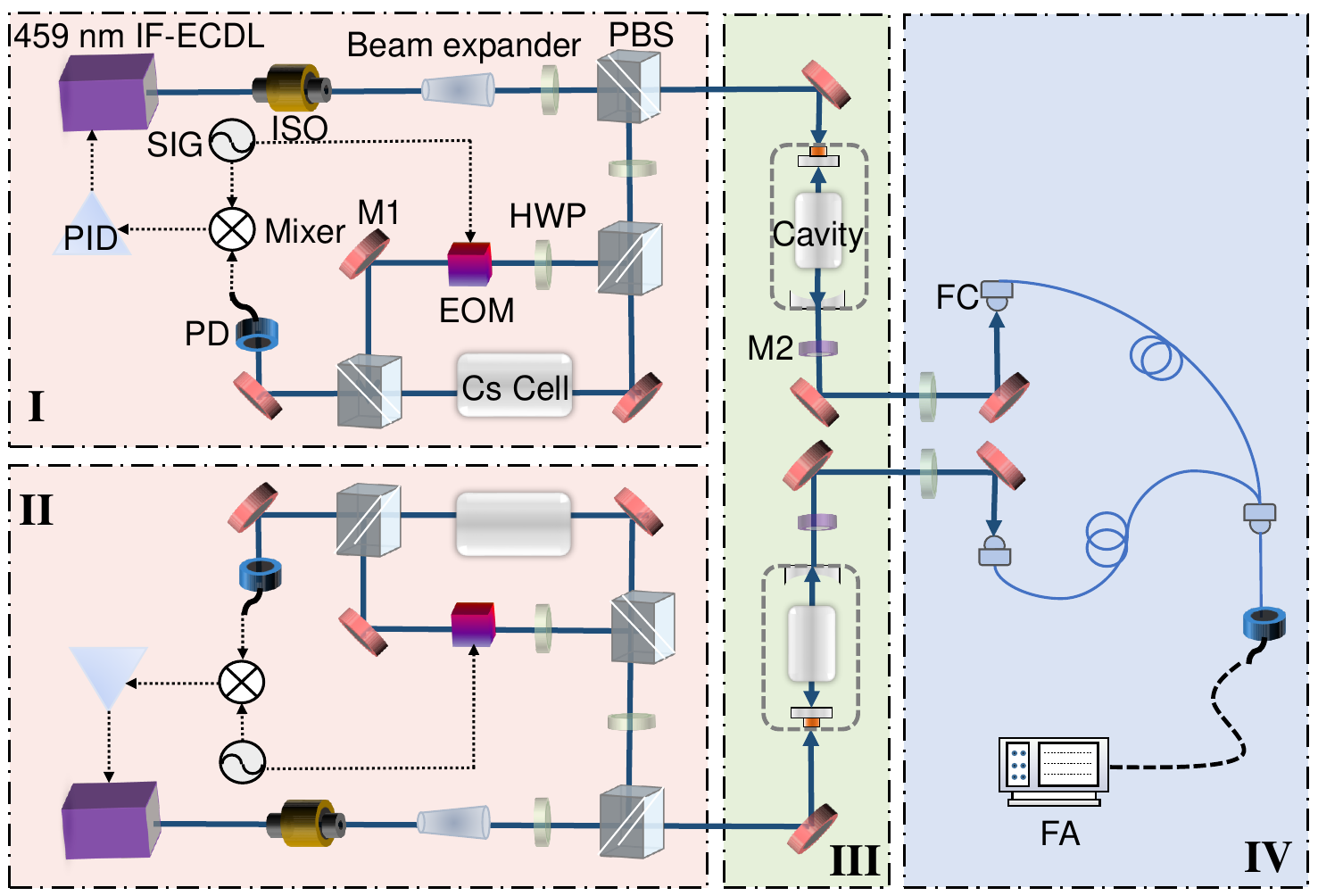}
	\caption{\label{fig5}
		\textbf{Heterodyne beat between two extremely bad-cavity lasers.} Module (Mod) $\mathrm{\Rmnum{1}}$ and $\mathrm{\Rmnum{2}}$ are 459-nm external cavity diode lasers stabilized through the modulation transfer spectroscopy. Mod $\mathrm{\Rmnum{3}}$ contains two 1470-nm lasers and the heterodyne measurement is implemented in Mod $\mathrm{\Rmnum{4}}$. IF-ECDL, interference filter configuration external cavity diode laser; ISO, isolator; PBS, polarizing beam splitter; HWP, half-wave plate; EOM, electro-optic modulator; PD, photodetector; SIG, signal generator; PID, proportional-integral-derivative locking system; FC, fiber coupler; FA, frequency analyzer. M1 is the 459-nm reflective mirror and M2 is a dichroic mirror with the anti-reflection coating at 1470~nm and the high-reflection coating at 459~nm. 
	}
\end{figure}

\textbf{Heterodyne beat between two 1470-nm extremely bad-cavity lasers.} To measure the frequency shift of the extremely bad-cavity laser, two identical 1470-nm laser systems are built, where one laser plays the reference role. The specific schematic is depicted in Fig.~\ref{fig5}, which contains four modules (Mod): Mod $\mathrm{\Rmnum{1}}$ and Mod $\mathrm{\Rmnum{2}}$ are two identical 459-nm pump lasers utilizing the modulation-transfer-spectroscopy frequency stabilization method. Mod $\mathrm{\Rmnum{3}}$ has two 1470-nm extremely bad-cavity lasers, where the finesse of optical cavities is adjustable. In each laser system, the plane mirror of the optical cavity (length $ L=17~\mathrm{cm}$) is high transmission coated at 459~nm and the concave output mirror has a radius of curvature of 18~cm. A piezoelectric ceramic is bonded to the plane mirror so as to precisely adjust the cavity length. Mod $\mathrm{\Rmnum{4}}$ is for heterodyne detection, where a frequency analyzer is used to evaluate the central frequency and spectral linewidth of the beating signal between two extremely bad-cavity lasers.

\textbf{Optical gain bandwidth.} The optical gain bandwidth is related to the pump power broadening and the Doppler broadening of the laser transition. The atomic $6\mathrm{S}_{1/2}$ - $7\mathrm{P}_{1/2}$ transition (pump line) has a natural linewidth of $\gamma_{\mathrm{pump}}=1.1~\mathrm{MHz}$ and the corresponding saturation intensity is evaluated as $I_{\mathrm{s}}=1.27~\mathrm{mW~cm^{-2}}$ \cite{heavens1961radiative}. The 459-nm pump laser used in experiment has a spot area of $S=1.69~\mathrm{mm^{2}}$. The pump light intensity reaches $I=308~\mathrm{mW~cm^{-2}}$ at the typical pump power of 5.2~mW. Thus, the saturation broadening of the pump line is given by $\Gamma_{\mathrm{pump}}=\gamma_{\mathrm{pump}}\sqrt{(1+I/I_{\mathrm{s}})}=17.2~\mathrm{MHz}$. According to the velocity-selective mechanism, only atoms with a velocity in the pump-beam direction less than $v_{\mathrm{D}}=\Gamma_{\mathrm{pump}}\times(459~\mathrm{nm})=7.9~\mathrm{m~s^{-1}}$ can be effectively pumped to $7\mathrm{P}_{1/2}$ and then decay to $7\mathrm{S}_{1/2}$. Thus, the Doppler broadening of the atomic $6\mathrm{P}_{3/2}$ - $7\mathrm{S}_{1/2}$ line (laser transition) reaches $\Gamma_{\mathrm{D}}=v_{\mathrm{D}}/(1470~\mathrm{nm})=5.36~\mathrm{MHz}$. Combining the Doppler broadening with the natural linewidth of the laser transition $\gamma_{\mathrm{gain}}=1.81~\mathrm{MHz}$, one obtains the optical gain bandwidth $\Gamma_{\mathrm{gain}}=\gamma_{\mathrm{gain}}+\Gamma_{\mathrm{D}}=7.17~\mathrm{MHz}$. Using Eq.~\ref{eq.1} to obtain the different finesses, the corresponding cavity-mode linewdith $\Gamma_{\mathrm{cavity}}=\mathrm{FSR}/\mathcal{F}$ are all much wider than the gain bandwidth.

\begin{table}[H]
	\caption{\label{table1}%
		Cavity reflectance $R$, finesse $\mathcal{F}$, cavity linewidth $\Gamma_{\mathrm{cavity}}$, and pulling coefficient $P$ in experiment.
	}
	\begin{ruledtabular}
		\begin{tabular}{lccccc}
			$R$&$\mathcal{F}$&$\mathrm{log}_{10}(\mathcal{F}-2)$&$\Gamma_{\mathrm{cavity}}$ (MHz)&$P$\\
			\colrule\rule{0pt}{1.2em}%
			0.485\%&2.01&-1.91&438.4&0.0148\\
			2.08\%&2.05&-1.27&429.5&0.0151\\
			3.45\%&2.09&-1.04&421.8&0.0154\\
			18.3\%&2.60&-0.22&339.5&0.0192\\
			22.4\%&2.78&-0.11&317.2&0.0205
		\end{tabular}
	\end{ruledtabular}
\end{table}

\begin{figure}[htb]
	\includegraphics[width=\linewidth]{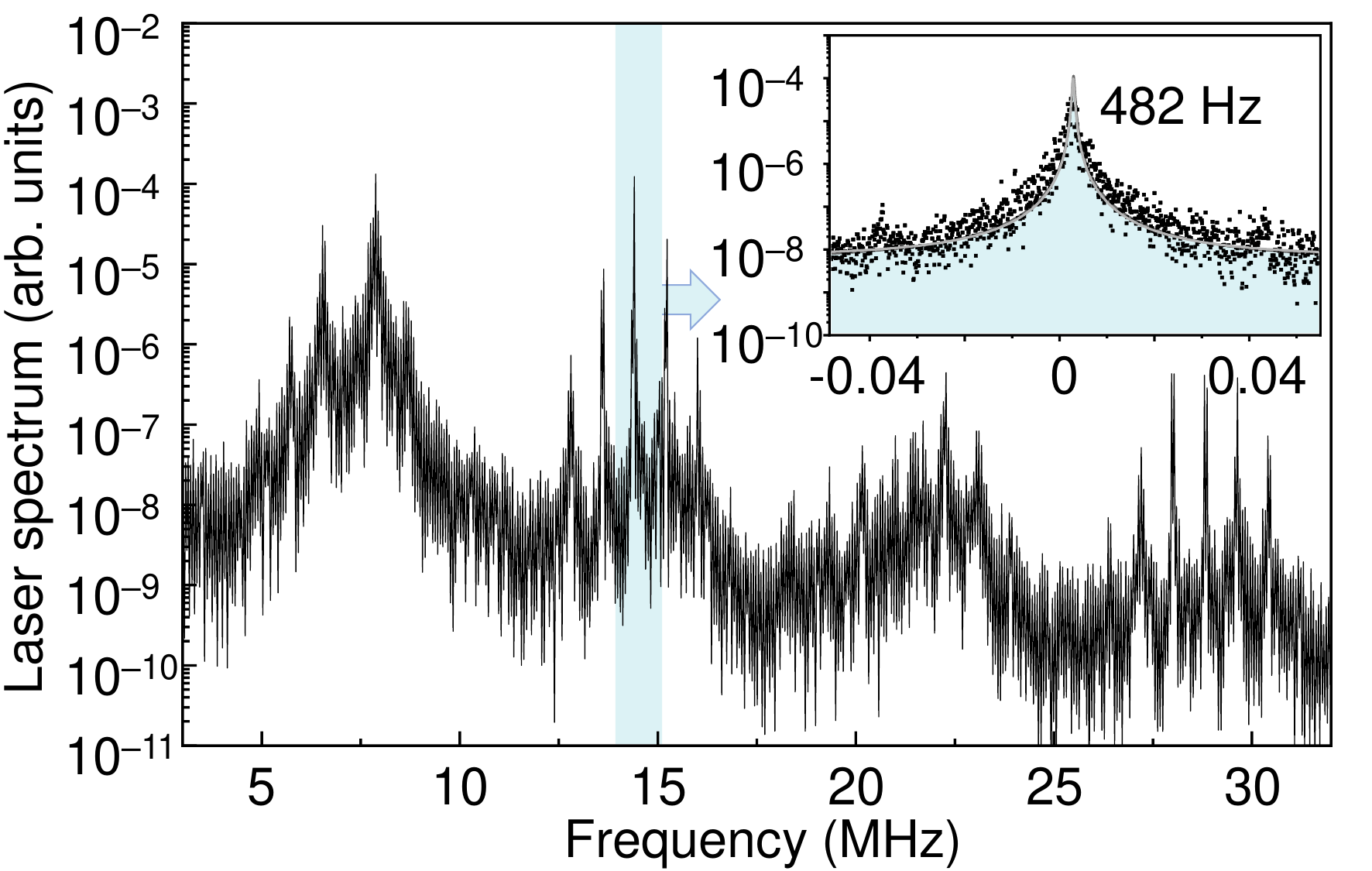}
	\caption{\label{fig6}
		\textbf{Beat spectrum between 1470-nm lasers based on different Zeeman sublevel transitions.} Insert symbols: experimental results with the resolution bandwidth (RBW) of 47~Hz. The Lorentz fit (solid line) results in a spectral linewidth of 482~Hz, indicating a linewidth of 341~Hz for each laser mode.
	}
\end{figure}

\textbf{Linewidth eliminating the common-mode noise.} 
We demonstrated the extremely bad-cavity laser under a weak magnetic field and measured the beating linewidth between Zeeman sublevels. By applying a weak magnetic field perpendicular to the direction of the light to the vapor cell, a splitting of the energy levels $7\mathrm{S}_{1/2}$ and $6\mathrm{P}_{3/2}$ occurs. As a result, beating signals between different Zeeman levels can be observed, as shown in Fig.~\ref{fig6}. Here, we select a typical beat frequency signal for Lorentz fitting, demonstrating a power spectrum with a linewidth of 482~Hz (see the Fig.~\ref{fig6} insert). Assuming equal contribution from each laser mode in the beating linewidth between two Zeeman sublevels, the linewidth of each laser mode is 341~Hz, primarily reflecting the limited linewidth of our extremely bad-cavity laser using thermal atoms. Because the two Zeeman sublevels share the same cavity, which has an elimination of the common-mode noise, such as the cavity length vibrations, the pumping power fluctuations, and the vapor cell temperature changes, the beating linewidth is narrower than that between two extremely bad-cavity lasers.

\begin{figure}[htp]
	\includegraphics[width=\linewidth]{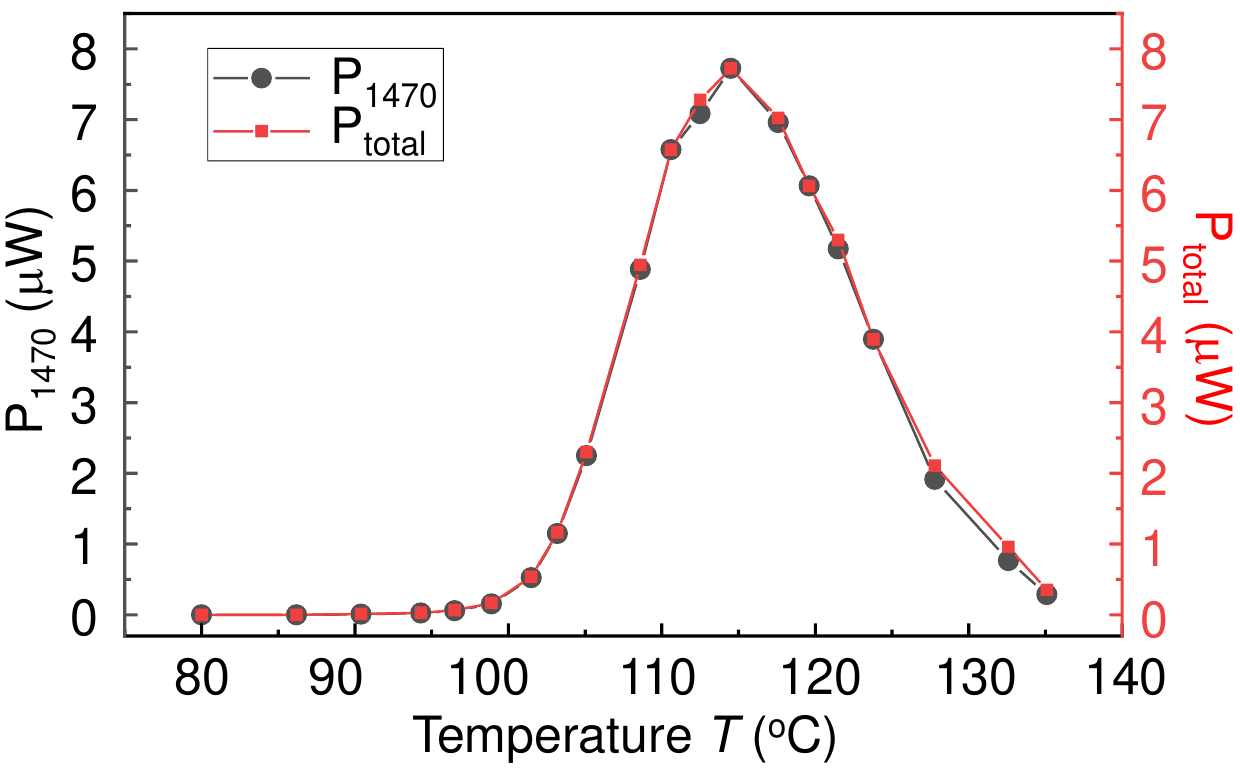}
	\caption{\label{fig7}
		\textbf{Mirrorless superradiance power vs. vapor cell temperature.} When measuring the 1470-nm mirrorless superradiance power (red squares), a long-pass interference filter is used to filter out the 1359-nm component in the output light. The total power of the output light (black circles) is measured in the absence of the long-pass interference filter. For all measurements, the 459-nm pump power is set at 5.2~mW.
	}
\end{figure}

\textbf{Mirrorless lasing.} The atoms in vapor cell (length $ l=5~\mathrm{cm}$) are directly pumped by the 459-nm laser in the absence of cavity mirrors. Cell windows are set at Brewster's angle to prevent any reflection of 1470-nm light. As shown in Fig.~\ref{fig7}, the 1470-nm laser power depends strongly on the vapor cell temperature $T$ and is maximized at $T=116.2^{\circ}\mathrm{C}$, where the atomic density $n = 5.2\times10^{13}~\mathrm{cm}^{-3}$ and the number of atoms reaches $N = n LS = 4.39\times10^{12}$. Due to the Doppler broadening, only thermal atoms with the velocity between $-v_{\mathrm{D}}/2$ and $v_{\mathrm{D}}/2$ can be effectively pumped to $7\mathrm{P}_{1/2}$. According to the Maxwell velocity distribution, the effective number of atoms contributing to the 1470-nm laser is then given by $N_{\mathrm{eff}} =N \int_{-v_{\mathrm{D}}/2}^{v_{\mathrm{D}}/2} \frac{1}{\sqrt{2\pi\Delta v}} \mathrm{e}^{(-v/\Delta v)^{2}} d v=7.55\times10^{10}$. Here, the velocity distribution width of thermal atoms is computed as $\Delta v=\sqrt{k_{\mathrm{B}}T/m}=156~\mathrm{{m~s^{-1}}}$($\gg v_{\mathrm{D}}$) with the atomic mass $m$ and Boltzmann constant $k_{\mathrm{B}}$. The mirrorless superradiance spectrum is measured through the beating signal with an extremely bad-cavity laser ($R=0.485\%$).

In principle, the population inversion can be also achieved on the $7\mathrm{S}_{1/2}$ - $6\mathrm{P}_{1/2}$ transition (wavelength of 1359~nm). However, the corresponding lasing action is inhibited in practice (see Methods). This is because the dipole moment of the $7\mathrm{S}_{1/2}$ - $6\mathrm{P}_{3/2}$ transition ($2.4ea_{0}$ with elementary charge $e$ and Bohr radius $a_{0}$) is larger than that of the $7\mathrm{S}_{1/2}$ - $6\mathrm{P}_{1/2}$ transition ($1.8ea_{0}$). Thus, more atoms in $7\mathrm{S}_{1/2}$ decay to the $6\mathrm{P}_{3/2}$ level, and eventually the competition prevents the lasing action upon the $6\mathrm{P}_{1/2}$ - $7\mathrm{S}_{1/2}$ transition. To prove that the lasing action on the $6\mathrm{P}_{1/2}$-$7\mathrm{S}_{1/2}$ transition (wavelength of 1359 nm) is inhibited, a long-pass interference filter at 1359~nm is used to filter out the 1359-nm component in the output light. It is found that the power of the filtered light is almost equal to that of the unfiltered light (see Fig.~\ref{fig7}). 

Moreover, within the pump power range, we did not observe the saturation behavior of the 1470-nm laser power, although the 459-nm pump beam intensity well exceeds the saturation intensity of the atomic $6\mathrm{S}_{1/2}$-$7\mathrm{P}_{1/2}$ transition. This is because a high pump power may excite more thermal atoms in different velocity groups from $6\mathrm{S}_{1/2}$ to $7\mathrm{P}_{1/2}$, thereby always enhancing the population inversion on the laser transition. The velocity distribution of effective atoms contributing the 1470-nm laser is far narrower than the Doppler velocity distribution.

\textbf{Data availability}
Data underlying the results of this study are available from the authors upon request.

\textbf{Acknowledgments}
This research was funded by the National Natural Science Foundation of China (NSFC) (91436210), Innovation Program for Quantum Science and Technology (2021ZD0303200), China Postdoctoral Science Foundation (BX2021020), and Wenzhou Major Science \& Technology Innovation Key Project (ZG2020046).

\textbf{Author contributions}
J.C. conceived the idea to use an extremely low-finesse cavity to realize the extremely bad-cavity laser as a stable active optical clock. J.Z., T.S. and J.M. performed the experiments. J.Z., D.Y. and T.S. carried out the theoretical calculations and wrote the manuscript. T.S., J.M. and J.C. provided revisions.

\textbf{Competing interests}
The authors declare no competing interests.

%\begin{acknowledgments}
%\end{acknowledgments}

\bibliography{bad-cavity}

\end{document}